\documentstyle[12pt,aps,epsfig]{revtex}

\def\be{\begin{equation}}
\def\ee{\end{equation}}
\def\L{\bbox{\cal L}}
\def\H{\bbox{\cal H}}

\def\Q{\bbox{\cal Q}}
\draft
\title{Bohmian Quantum Gravity in the Linear Field Approximation}
\author{ALI SHOJAI\thanks{Email: SHOJAI@IPM.IR}}
\address{Physics Department, Tehran University, End of North Karegar St., Tehran 14352, IRAN}
\address{and}
\address{Institute for Studies in Theoretical Physics and Mathematics, P.O.Box 19395-5531, Tehran,
IRAN}
\author{FATIMAH SHOJAI\thanks{Email: FATIMAH@IPM.IR}}
\address{Physics Department, Iran University of Science and Technology, P.O.Box 16765--163, Narmak, Tehran,
IRAN}
\address{and}
\address{Institute for Studies in Theoretical Physics and Mathematics, P.O.Box 19395-5531, Tehran,
IRAN}
\begin{document}
\maketitle
\begin{abstract}
In this paper we have applied Bohmian quantum theory to the linear
field approximation of gravity and present a self--consistent
quantum gravity theory in the linear field approximation. The
theory is then applied to some specific problems, the Newtonian
limit, and the static spherically symmetric solution. Some
observable effects of the theory are investigated.
\end{abstract}
\section{INTRODUCTION}
There are several approaches to the quantization of gravity. Some
important ones are, the canonical formalism (or the WDW
formalism)\cite{WDW}, Narlikar and Padmanabhan approach to
quantization of the conformal degree of freedom of the space--time
metric\cite{NAR}, the standard Bohmian quantum
gravity\cite{HOL,HOR}, and a new approach\cite{GEO,CON,STQ}
relating quantum mechanics and the geometry of the space--time.
Each of the above theories have some positive and some negative
aspects, so that there is no universally accepted theory of
quantum gravity. As some authors have mentioned, Bohmian quantum gravity has some noticable aspects.
It has not
the problem of time, the measurement problem and the difficulty
with the meaning of the wavefunction\cite{PRO}. Also it presents a
trajectory for the system just like in the classical
domain\cite{HOL}.

To understand how this theory works, let us briefly explain the
Bohmian quantum mechanics of a single non--relativistic
particle\cite{BOH}. In the standard quantum mechanics, the
particle's wavefunction satisfies the Schr\"{o}dinger equation:
\be
i\hbar\frac{\partial \Psi}{\partial t}=-\frac{\hbar^2}{2m}
\nabla^2\Psi+V(\overrightarrow{x};t)\Psi \ee If one decomposes the
wavefunction into its norm and its phase:
\be
\Psi=R\exp(iS/\hbar)
\ee
one gets the following two equations:
\be
\frac{\partial S}{\partial t}+\frac{|\overrightarrow{\nabla}S|^2}{2m}
+V+\Q=0
\ee
\be
\frac{\partial R^2}{\partial t}+\overrightarrow{\nabla}\cdot
\left(R^2\frac{\overrightarrow{\nabla}S}{m}\right)=0
\ee
where
\be
\Q=-\frac{\hbar^2}{2m}\frac{\nabla^2R}{R}
\ee
As it can be seen, we have a modified Hamilton--Jacobi equation
and the continuity equation provided we assume that the particle's
trajectory is given by the classical relation:
\be
\overrightarrow{v}=\frac{\overrightarrow{\nabla}S}{m}
\label{gui}
\ee
The Newton's equation of motion can be derived by taking the
gradiant of the Hamilton--Jacobi equation and using the guidance
relation (\ref{gui}). The result is:
\be
m\frac{d^2\overrightarrow{x}}{dt^2}=
-\overrightarrow{\nabla}V-\overrightarrow{\nabla}\Q \ee So the
Bohmian interpretation is as follows. To quantize any classical
system add a {\em quantum potential} to the classical
Hamilton--Jacobi equation. The quantum potential is given in terms
of the density of an hypothetical ensemble of the system ($R^2$)
and thus add the continuity equation to have a self--consistent
system of equations. 

Bohmian quantum gravity is also constructed successfully\cite{HOR,HOL}.
But it has also some problems. The most essential one is perhaps
the lack of general covariance\cite{COV}. Although some authors
look at this property as an essential character of quantum
gravity\cite{STA}.

Investigating Bohmian quantum gravity in the linear field
approximation of gravity is perhaps a good idea, because
as it is the case for classical gravity some important effects can
be understood better, in the linear approximation. In this paper we shall do this and then study the theory for the Newtonian limit and the static spherically symmetric metric, and some observable phenomena are predicted.

\section{LINEAR FIELD BOHMIAN QUANTUM GRAVITY}
Our plan in this section is to make Bohmian quantum gravity in the
linear field approximation, i.e. when the space--time metric is
expanded about the flat Minkowski space--time as:
\be
g_{\mu\nu}=\eta_{\mu\nu}+h_{\mu\nu} \ee and considering only
linear terms in $h_{\mu\nu}$. The lagrangian density of the linear
field classical gravity is given by\cite{OHA}:
\be
\L=\frac{1}{2}\partial_\lambda\phi_{\mu\nu}\partial^\lambda\phi^{\mu\nu}
\ee
where as it is convenient we introduced the new field variables:
\be
\phi_{\mu\nu}=h_{\mu\nu}-\frac{1}{2}h\eta_{\mu\nu}
\ee 
The canonical momenta are defined as:
\be
\bbox{\Pi}_{\mu\nu}=\frac{\delta\L}{\delta\dot{\phi}^{\mu\nu}}
=\frac{\partial\phi_{\mu\nu}}{\partial t} \ee
and the hamiltonian
density is:
\be
\H=\frac{1}{2}\bbox{\Pi}_{\mu\nu}\bbox{\Pi}^{\mu\nu}+\frac{1}{2}
\nabla_i\phi_{\mu\nu}
\nabla^i\phi^{\mu\nu} \ee
 The general coordinate
transformation invariance of the linear field gravity is
guaranteed by choosing some specific gauge. The common gauge is
the harmonic gauge:
\be
\partial^\mu\phi_{\mu\nu}=0
\label{a}
\ee

The transition to the quantum theory can be achieved via Dirac's
canonical quantization scheme, in which we replace:
\be
\bbox{\Pi}_{\mu\nu}\rightarrow -i\hbar\frac{\delta}{\delta\phi_{\mu\nu}}
\ee
leading to the following wave equation:
\be
i\hbar\frac{\partial\Psi}{\partial t}=\left (-\frac{\hbar^2}{2}
\int d^3x' \frac{\delta^2}{\delta\phi_{\mu\nu}(x')\delta\phi^{\mu\nu}(x')}
+\frac{1}{2}\int d^3x' \nabla'_i\phi_{\mu\nu}
\nabla'^i\phi^{\mu\nu} \right )\Psi \ee where
$\Psi[\phi_{\alpha\beta}]$ is the wavefunctional.

The Bohmian interpretation of this wave equation can be achieved
via setting:
\be
\Psi=R\exp(iS/\hbar)
\ee
leading to the Hamilton--Jacobi equation:
\be
\frac{\partial S}{\partial t}+\frac{1}{2}\int
d^3x' \frac{\delta
S}{\delta\phi_{\mu\nu}(x')}\frac{\delta
S}{\delta\phi^{\mu\nu}(x')} +\frac{1}{2}\int d^3x'
\nabla'_i\phi_{\mu\nu}(x')
\nabla'^i\phi^{\mu\nu}(x')+\Q=0 \label{HJ} \ee and
the continuity equation:
\be
\frac{\partial R^2}{\partial t}+\int
d^3x'
\frac{\delta}{\delta\phi_{\mu\nu}(x')}\left ( R^2\frac{\delta
S}{\delta\phi^{\mu\nu}(x')}\right )=0 \label{CR} \ee with
the quantum potential defined as:
\be
\Q=-\frac{\hbar^2}{2R}\int
d^3x'
\frac{\delta^2R}{\delta\phi_{\mu\nu}(x')\delta\phi^{\mu\nu}(x')}
\label{QP} \ee The Bohmian trajectories can be obtained via the
guidance equation:
\be
\frac{\partial\phi_{\mu\nu}}{\partial t}=\frac{\delta
S}{\delta\phi^{\mu\nu}} \label{GE} \ee and the gauge condition is
given by the equation (\ref{a}).

It is important here to note that we have not
included the gauge condition as it is always done. That is to say
we have not set:
\be
\partial_\mu\phi^{\mu\nu}=0\Rightarrow\left
(\Pi^{0\nu}+\partial_i\phi^{i\nu}\right )\left|\Psi\right\rangle=
\left
(-i\hbar\frac{\delta}{\delta\phi_{0\nu}}+\partial_i\phi^{i\nu}\right
)\left|\Psi\right\rangle=0 \ee 
Because this forces the state to satisfy the gauge condition and the Bohmian trajectories are not necessarily consistent with it.
If one wishes to save the gauge condition at the level of trajectories, one should apply it on the Bohmian
trajectories $\phi^{\mu\nu}(\overrightarrow{x},t)$.

The field equation can be derived by taking variation of the
Hamilton--Jacobi equation (\ref{HJ}) with respect to
$\phi_{\mu\nu}$. The result is:
\be
\Box\phi_{\mu\nu}(x)=-\frac{\delta\Q}{\delta\phi^{\mu\nu}(x)}
\ee 

Let us now, apply these results to the Newtonian
approximation. In this limit the metric is given by:
\be
h_{00}=h=2\Phi;\ \ \ \ \ h_{ij}=h_{0i}=0
\ee
where $\Phi$ is the
Newton's gravitational potential. The $\phi$--field is given by:
\be
\phi_{00}=\Phi;\ \ \ \ \ \ \phi_{ij}=\Phi\delta_{ij};\ \ \ \ \ \
\phi_{0i}=0
\ee
and we have the identities:
\be
\frac{\delta}{\delta\phi_{00}}=\frac{\delta}{\delta\Phi}; \ \ \ \
\ \ \
\frac{\delta}{\delta\phi_{ij}}=\delta_{ij}\frac{\delta}{\delta\Phi}
\ee
So the Hamilton--Jacobi equation and the continuity equation are given by:
\be
\frac{\partial S}{\partial t}+\int d^3x'\left ( \frac{\delta S}{\delta\Phi(x')}\right )^2 +\frac{1}{2}\int d^3x'
\nabla'_i\Phi(x')
\nabla'^i\Phi(x') +\Q=0 \ee
\be
\frac{\partial R^2}{\partial t}+4\int
d^3x' \frac{\delta}{\delta\Phi(x')}\left (
R^2\frac{\delta S}{\delta\Phi(x')}\right )=0
\ee
where the quantum potential is given by:
\be
\Q=-\frac{2\hbar^2}{R}\int d^3x'\frac{\delta^2
R}{\delta\Phi(x')^2} \ee The guidance relation is:
\be
\frac{\partial\Phi}{\partial t}=\frac{\delta S}{\delta\Phi} \ee
and the condition of the Newtonian approximation is given by:
\be
\frac{\partial\Phi}{\partial t}=0 \ee instead of the relation
(\ref{a}).
As a result of this last equation we have:
\be
\frac{\partial\Phi}{\partial t}=\frac{\delta S}{\delta\Phi}=0 \ee
so we have encountered with a pure quantum case\footnote{That is a
case in which quantum potential balances the classical potential.
So the quantum potential is not ignorable and we have no classical
limit\cite{HOL}.} and so we have:
\be
-E+\frac{1}{2}\int d^3x'\nabla'_i\Phi(x')
\nabla'^i\Phi(x')=\Q \ee where we have assumed
$\partial S/\partial t=-E$, because in the Newtonian limit
physical quantities should not depend on time. The continuity equation is satisfied identically.

The field equation can be derived by differentiating the
Hamilton--Jacobi equation with respect to $\Phi$. The result is:
\be
\nabla^2\Phi=\frac{\delta\Q}{\delta\Phi(x)} \ee Until now we
have encountered with the free linear field Bohmian quantum
gravity. That is we have not considered the matter. If one wants
to add the matter field semi--classically\footnote{In order to
take into account the quantum effects of matter, one should start
from the linear field lagrangian density with matter, and then
proceed as in the previous section. The only difference would be
that the matter quantum potential is also present in the field
equations. We did not do so here, because in this work, the quantum effects of gravity are of consider.}, one should add merely
the source term $-\kappa\rho$ to the above equation, where $\rho$
is the matter density and $\kappa$ is the gravitational constant.
So we have:
\be
\nabla^2\Phi=-\kappa\rho+ \frac{\delta\Q}{\delta\Phi(x)} \label{NQG}\ee 
\section{STATIC SPHERICALLY SYMMETRIC SOLUTION}
Let's now look for static spherically symmetric solution of the equations of the previous section. We do this in the case of Newtonian and weak field approximations.
\begin{itemize}
\item[(a)] First we work in the Newtonian approximation.
Let's choose $R$ as a packet around the classical solution of
the gravitational field equation:
\be
R=\exp\left\{-\frac{\alpha}{2}\left [\Phi(x)-\Phi^{(c)}(x)\right
]^2\right \} \label{WP} \ee where $\Phi^{(c)}$ is the classical
field, i.e. the field satisfying the equation (\ref{NQG}) without
$\Q$. This leads to the following quantum potential:
\be
\Q=-\alpha\hbar^2\left [ 1-\alpha(\Phi-\Phi^{(c)})^2\right ] \ee
so the field equation is given by:
\be
\nabla^2\Phi=-\kappa\rho-\alpha^2\hbar^2(\Phi-\Phi^{(c)})
\label{TTT}\ee Let us look for the gravitational potential, when
we have a point--like source:
\be
\rho=M\delta(\vec{x}) \ee We solve the equation (\ref{TTT}) by
setting:
\be
\Phi=\Phi^{(c)}+f=-\frac{\kappa M}{r}+f \ee so we have:
\be
\nabla^2f=-\alpha^2\hbar^2f \ee 
The solution is given by:
\be
\Phi=-\frac{\kappa M}{r}+\sum_{lm}\left [ a_{lm}\ j_l(\alpha\hbar
r)+b_{lm}\ n_l(\alpha\hbar r)\right ] Y_{lm}(\theta,\varphi) \ee
where $a_{lm}$ and $b_{lm}$ are constants.

Some points about this solution is of interest:
\begin{itemize}
\item Because of the asymptotic behaviour of the spherical Bessel
functions:
\be
j_l(x)\sim\frac{1}{x}\sin\left (x-\frac{l\pi}{2}\right )\ \ \
\text{as}\ \ x\rightarrow\infty \ee
\be
n_l(x)\sim\frac{1}{x}\cos\left (x-\frac{l\pi}{2}\right )\ \ \
\text{as}\ \ x\rightarrow\infty \ee one can see that this quantum
solution also behaves as $\frac{1}{r}$ at large distances from the
source.
\item It must be noted that the weak field approximation is only applicable for large $r$'s. So this solution does not tell anything about the event horizon or about the singularity.
\item If one sets $M=0$, so that we have no source at all, we have
a pure quantum solution in which we have source--less gravity. This means that the quantum fluctuations of gravity can produce observable gravity.
\item Because of the oscillatory nature of the spherical Bessel
functions, the gravitational potential is also oscillatory. So we
have a set of periodic stable points, where the potential is
minimum, and a test particle can be at rest at that locations
(This is true also for the case $M=0$). It must be noted that, if we
choose a sharp wave packet ($\alpha\rightarrow\infty$ in equation
(\ref{WP})), the frequency of such oscillations is very large.
\end{itemize}

\item[(b)] Let's now investigate the static spherically
symmetric metric in the linear field approximation. The relations (\ref{HJ}) and (\ref{CR}) would be
now:
\be
\frac{1}{2}\int d^3x'
\frac{\delta S}{\delta\phi_{\mu\nu}(x')}\frac{\delta
S}{\delta\phi^{\mu\nu}(x')} +\frac{1}{2}\int d^3x'
\nabla'_i\phi_{\mu\nu}(x')
\nabla'^i\phi^{\mu\nu}(x')+\Q=0 \ee and
\be
\int d^3x'
\frac{\delta}{\delta\phi_{\mu\nu}(x')}\left (
R^2\frac{\delta S}{\delta\phi^{\mu\nu}(x')}\right )=0
\ee
The Bohmian trajectories can be obtained via the guidance
equation:
\be
0=\frac{\partial\phi_{\mu\nu}}{\partial t}=\frac{\delta
S}{\delta\phi_{\mu\nu}}
\ee
and the gauge condition (\ref{a}).
It is a simple task to see that the corrected field equation is
given by:
\be
\nabla^2\phi_{\mu\nu}= \frac{\delta\Q}{\delta\phi^{\mu\nu}(x)}
 \label{x2} \ee As in the
Newtonian case, choosing:
\be
R=\exp\left \{-\frac{\alpha}{2}\left [
\phi_{\mu\nu}-\phi^{(c)}_{\mu\nu}\right ]\left [
\phi^{\mu\nu}-\phi^{(c)\mu\nu}\right ]\right \} \ee we have
the following relation for the quantum potential:
\be
\Q=\frac{\alpha\hbar^2}{2}\left \{4-\alpha\left [
\phi_{\mu\nu}-\phi^{(c)}_{\mu\nu}\right ]\left [
\phi^{\mu\nu}-\phi^{(c)\mu\nu}\right ]\right \} \ee so the
field equations are:
\be
\nabla^2\phi_{\mu\nu}+\alpha^2\hbar^2\left
[\phi_{\mu\nu}-\phi^{(c)}_{\mu\nu}\right ]=0;\ \ \ \ \
\partial_\mu\phi^{\mu\nu}=0
\ee
writing:
\be
\phi_{\mu\nu}=\phi^{(c)}_{\mu\nu}+f_{\mu\nu}=\frac{C_{\mu\nu}}{r}+f_{\mu\nu}
\ee where $\phi^{(c)}_{\mu\nu}=\frac{C_{\mu\nu}}{r}$ with
$C_{\mu\nu}$ constants, is the classical solution\cite{OHA}.  The
field equation leads to:
\be
\nabla^2f_{\mu\nu}+\alpha^2\hbar^2f_{\mu\nu}=0 \ee The above
equation can be solved just as in the previous section. Then one
must apply the gauge condition. This would leads to some relation
for constants $C_{\mu\nu}$. The solution is:
\be
\phi_{\mu\nu}=\left (
\begin{tabular}{cccc}
$-\frac{2r_s}{r}+f$&0&0&0\\ 0&0&0&0\\ 0&0&0&0\\ 0&0&0&0
\end{tabular}
\right ) \ee where $f(r,\theta,\varphi)$ is
just as in the previous section, i.e.:
\be
f(r,\theta,\varphi)=\sum_{lm}\left [ a_{lm}\ j_l(\alpha\hbar
r)+b_{lm}\ n_l(\alpha\hbar r)\right ] Y_{lm}(\theta,\varphi) \ee
and $r_s$, the only non vanishing element of $C_{\mu\nu}$, is the
Schwarzschild  radius. This solution leads to:
\be
h_{00}=-\frac{r_s}{r}+\frac{f}{2}
\ee
\be
h_{0i}=0
\ee
\be
h_{ii}=-\frac{r_s}{r}+\frac{f}{2} \ee So the metric
$g_{\mu\nu}=\eta_{\mu\nu}+h_{\mu\nu}$ is given by:
\be
g_{\mu\nu}=\left (
\begin{tabular}{cccc}
$1-\frac{r_s}{r}+\frac{f}{2}$&0&0&0\\
0&$-1-\frac{r_s}{r}+\frac{f}{2}$&0&0\\
0&0&$-1-\frac{r_s}{r}+\frac{f}{2}$&0\\
0&0&0&$-1-\frac{r_s}{r}+\frac{f}{2}$
\end{tabular}
\right ) \ee
\end{itemize}
\section{SOME OBSERVABLE RESULTS}
\begin{figure}
\unitlength 1.00mm
\linethickness{1pt}
\begin{center}
\begin{picture}(103.33,106.33)
\put(10.00,100.00){\makebox(0,0)[cc]{$\bbox{\uparrow}$}}
\put(10.00,10.00){\line(0,1){90.00}}
\put(100.00,9.80){\makebox(0,0)[cc]{$\bbox{\rightarrow}$}}
\put(10.00,10.00){\line(1,0){90.00}}
\put(103.33,10.00){\makebox(0,0)[cc]{$b$}}
\put(10.00,106.33){\makebox(0,0)[cc]{$\theta$}}
\bezier{520}(15.00,89.00)(29.00,21.67)(89.67,17.00)
\bezier{80}(15.00,88.33)(14.33,73.33)(19.33,73.00)
\bezier{76}(19.33,73.00)(26.00,71.67)(25.00,59.67)
\bezier{76}(25.00,59.67)(23.67,49.00)(32.33,46.67)
\bezier{72}(32.33,46.67)(42.00,44.67)(41.00,37.33)
\bezier{80}(41.00,37.33)(44.33,25.67)(52.33,28.00)
\bezier{68}(52.33,28.00)(62.00,28.67)(65.33,22.33)
\bezier{56}(65.33,22.33)(70.67,16.33)(78.00,18.00)
\bezier{44}(78.00,18.00)(85.33,19.67)(90.00,17.33)
\put(23.67,82.00){\makebox(0,0)[ll]{classical prediction}}
\put(19.00,78.00){\makebox(0,0)[ll]{$\swarrow$}}
\put(39.33,46.00){\makebox(0,0)[ll]{$\swarrow$}}
\put(44.33,49.00){\makebox(0,0)[ll]{quantum prediction}}
\end{picture}
\end{center}
\caption{Classical and quantum light deflection in terms of the
impact parameter}
\label{fig}
\end{figure}
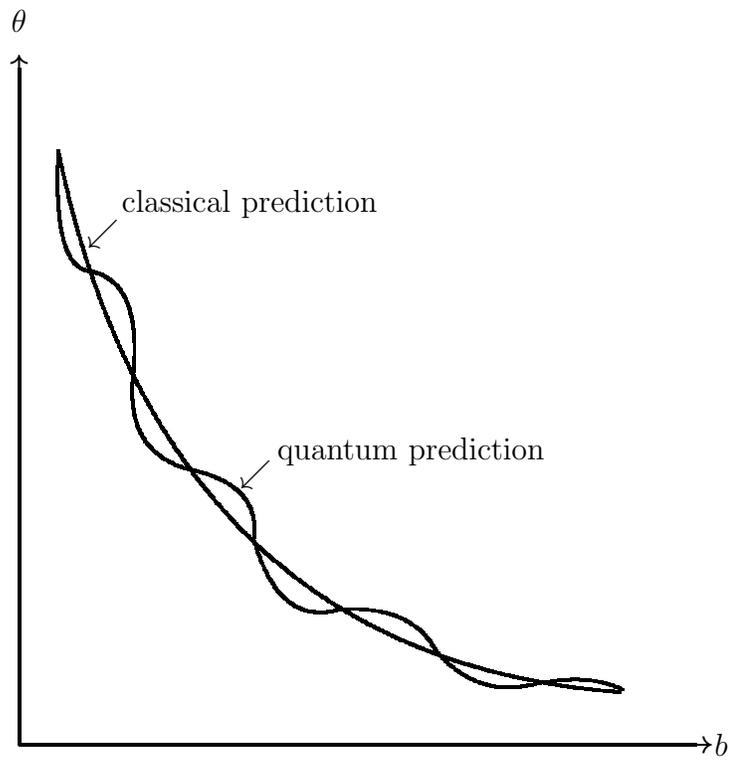

\begin{figure}
\vspace{1.2in}
\begin{center}
\epsffile{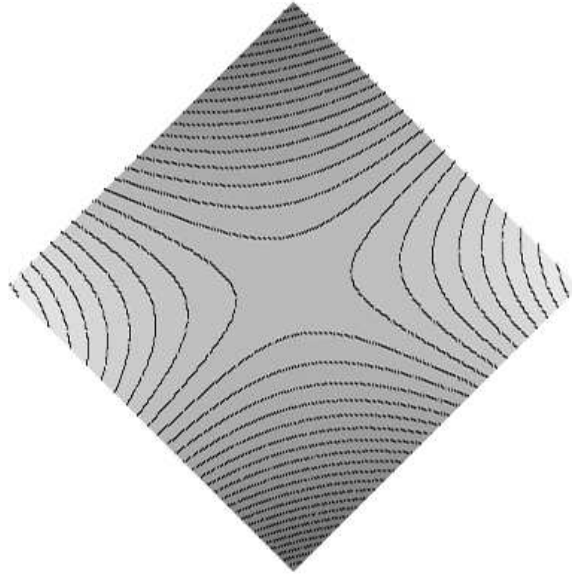}
\end{center}
\caption{Classical tidal force contours} \label{cc}
\end{figure}

\begin{figure}
\vspace{1.2in}
\begin{center}
\epsffile{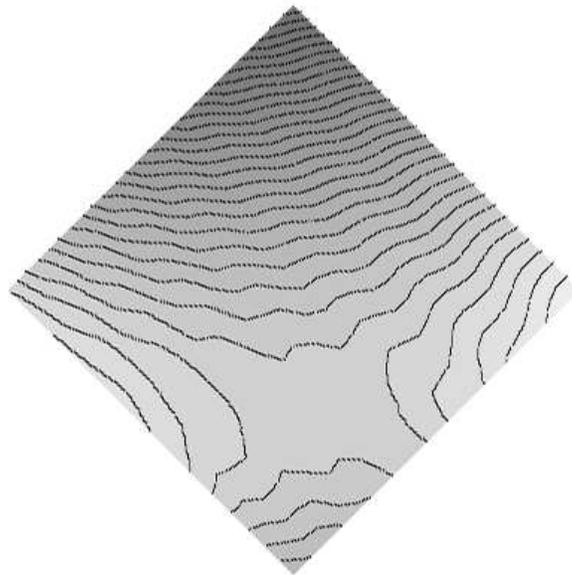}
\end{center}
\caption{Quantum tidal force contours} \label{cq}
\end{figure}

\begin{figure}
\vspace{1.2in}
\begin{center}
\epsffile{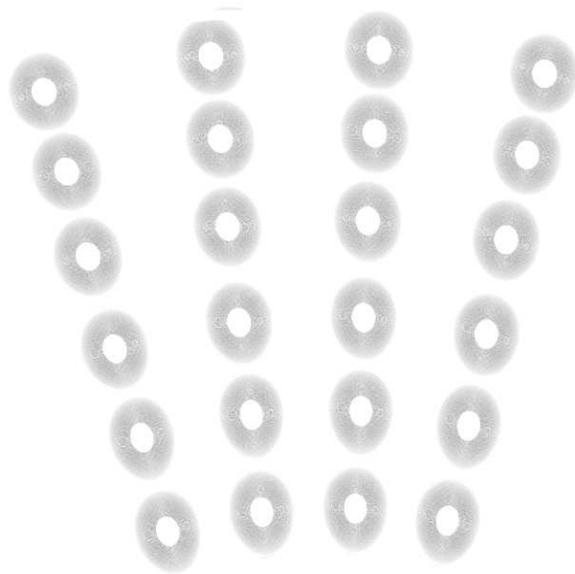}
\end{center}
\caption{Classical shape of a freely falling sphere} \label{sc}
\end{figure}

\begin{figure}
\vspace{1.2in}
\begin{center}
\epsffile{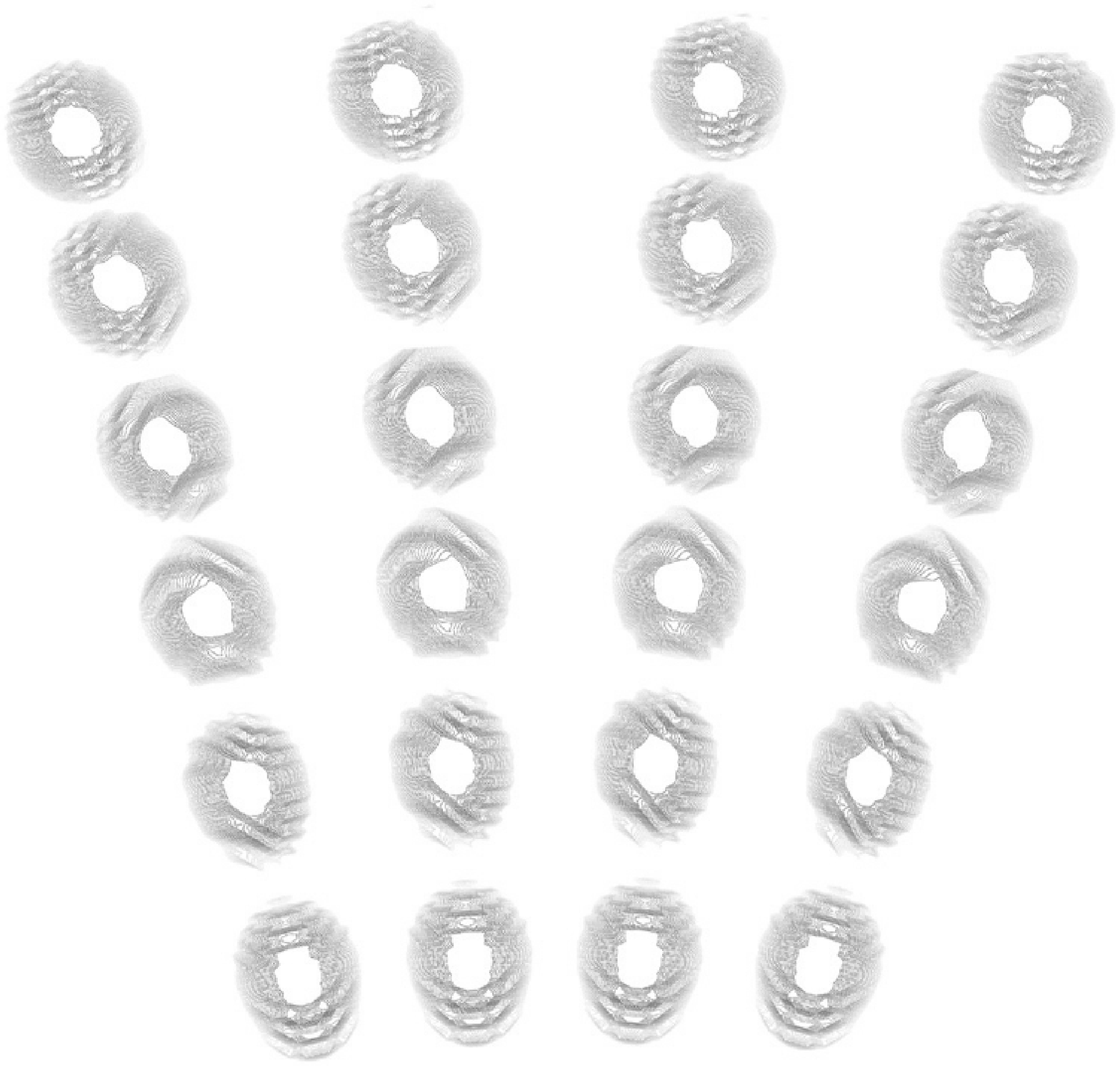}
\end{center}
\caption{Quantum shape of a freely falling sphere} \label{sq}
\end{figure}

At this end let us look for some observable results of the static spherical symmetric 
solution of the previous section. It is a good idea to see that the extra quantum terms
how affect the classical results of the Schwarzschild metric. We present here two effrcts.
\begin{enumerate}
\item Light deflection

As
an example consider the light bending effect. Suppose a light ray
approaches a star from infinity along the $z$ direction and in the $x-z$ plane with impact parameter $b$. The light ray path would be given
by the null geodesic relation, written in the linear field
approximation\cite{OHA}:
\be
dk_\mu-\frac{1}{2}\left (\partial_\mu h^\alpha_\beta\right )
k_\alpha dx^\beta=0 \ee where $k_\mu$ is the wave number. This
equation simplifies to:
\be
dk_x=-\frac{\omega}{2}\frac{\partial}{\partial x}\left (
-\frac{2r_s}{r}+f\right )_{x=b}dz \label{chom}\ee where $\omega$
is the frequency of the light beam.

For simplicity, we assume only the term $a_{00}$ contribute in $f$
so we set\footnote{If one wants to have a metric not dependent on the spherical angles at infinity, only the coefficients of $Y_{00}$ are nonzero. The behaviour of $j_0$ and $n_0$ are very similar, so for simplicity we only consider the first one. Considering the second does not change the physics.}:
\be
f=a_{00}j_0(\alpha\hbar r)=a_{00}\frac{\sin\alpha\hbar r}{\alpha\hbar r} \ee
Integrating the relation (\ref{chom}), one gets:
\be
\theta\simeq\frac{\int_{-\infty}^{+\infty}dk_x}{k_z}=-\frac{2r_s}{b}-a_{00}\left
( \cos\alpha\hbar b-\frac{\sin\alpha\hbar b}{\alpha\hbar b}\right
) \ee 
in which $\theta$ is the angle of bending.
The first term represents the classical light bending
relation, while the second term is the quantum corrections. The
dependence of the bending angle on the impact parameter is plotted
in the figure (\ref{fig}).

As it can be seen, the quantum solution is an oscillating curve
around the classical curve. The amplitude of the oscillations is
scaled by $a_{00}$, which is an arbitrary constant. So if any
oscillatory behaviour be observed in the light bending effect, it
can be described by the quantum effects.
\item Tidal forces

In order to investigate how the quantum effects change the curvature of the space-time, it is a good idea to derive the tidal forces. This is a simple task, noting that the tidal forces are in fact the gradient in the gravitational field in locally freely falling frame.
In figures (\ref{cc}) and (\ref{cq}) the classical and quantum fieldplots of tidal forces at some point are ploted. Also to see the effect better, in figures (\ref{sc}) and (\ref{sq}) the classical and quantum shape of a sphere falling on a spherically symmetric mass distribution is shown. Note the quantum oscillations in the shape in the last figure.
\end{enumerate}
{\bf Acknowledgement} The first author wishes to thank Tehran University for supporting this project under the grants provided by the research council.

\epsfxsize=3in \epsfysize=3in
\epsfxsize=3in \epsfysize=3in
\epsfxsize=3in \epsfysize=3in
\epsfxsize=3in \epsfysize=3in

\begin{thebibliography}{99}
\bibitem{WDW}
B.S. De Witt, {\it Phys. Rev.\/}, {\bf 160}, 1113, 1967;\\ J.A.
Wheeler, in {\it Batelle Remcomfres\/}, eds. C. De witt and J.A.
Wheeler (Benjamin, New York, 1968).
\bibitem{NAR}
J.V. Narlikar and T. Padmanabhan, {\it Phys. Rep.\/} {\bf 100},
151, 1983;\\ T. Padmanabhan, {\it Gravitation, Gauge Theories and
the Early Universe\/}, eds. B.R. Iyer et.al., 373, (Kluwer, 1989).
\bibitem{HOL}
P.R. Holland, {\it The Quantum Theory of Motion\/}, (Cambridge
Univ. Press, 1993).
\bibitem{HOR}
T. Horiguchi, {\it Mod. Phys. Lett. A.\/}, {\bf 9}, 16, 1429,
1994.
\bibitem{GEO}
F. Shojai and M. Golshani, {\it Int. J. Mod. Phys. A.\/}, 13, 4,
677, 1998.
\bibitem{CON}
F. Shojai, A. Shojai and M. Golshani, {\it Mod. Phys. Lett. A.\/},
13, 34, 2725, 1998.
\bibitem{STQ}
F. Shojai, A. Shojai and M. Golshani, {\it Mod. Phys. Lett. A.\/},
13, 36, 2915, 1998.
\bibitem{PRO}
See e.g. C. Kiefer, Lanl report gr-qc/9906100, in the Proceedings
of the Karpacz Winter School on {\it From Cosmology to Quantum
Gravity\/}, Springer, 1999;\\ J.B. Hartle, Quantum Cosmology:
Problems for the 21st century, in {\it Physics in the 21st
century\/}, edited by K. Kikkawa, H. Kunimoto, H. Ohtsubo, World
Scientific, Singapore, 1997.
\bibitem{BOH}
D. Bohm, {\it Phys. Rev.\/}, {\bf 85}, 166, 1952;\\ D. Bohm, {\it
Phys. Rev.\/}, {\bf 85}, 180, 1952;\\ D. Bohm and B.J. Hiley, {\it
The Undivided Universe\/}, Routledge, 1993.
\bibitem{COV}
F. Shojai and M. Golshani, {\it Int. J. Mod. Phys. A.\/}, 13, 13,
2135, 1998.
\bibitem{STA}
V. Shtanov, {\it Phys. Rev.\/} {\bf D54}, 4, 1996.
\bibitem{OHA}
S. Weinberg,{\it Gravitation and Cosmology\/}, (Wiley, New York,
1972);\\ H.C. Ohanian et.al., {\it Gravitation and Space--time\/},
2nd edition, WW Norton \& Co., 1994.
\bibitem{THE}
A. Shojai, PhD Thesis, Sharif University of Technology, 1997.
\end{thebibliography}
\end{document}